# Generating synthetic star catalogs from simulated data for next-gen observatories with py-ananke

Adrien C. R. Thob [1,¶], Robyn E. Sanderson [1], Andrew P. Eden [2], Farnik Nikakhtar [3], Nondh Panithanpaisal [1,4,5], Nicolás Garavito-Camargo [6], and Sanjib Sharma [7]

**1** Department of Physics & Astronomy, University of Pennsylvania, 209 S 33rd Street, Philadelphia, PA 19104, USA **2** Department of Aerospace, Physics and Space Sciences, Florida Institute of Technology, Melbourne, FL 32901, USA **3** Department of Physics, Yale University, New Haven, CT 06511, USA **4** The Observatories of the Carnegie Institution for Science, 813 Santa Barbara Street, Pasadena, CA 91101, USA **5** TAPIR, Mailcode 350-17, California Institute of Technology, Pasadena, CA 91125, USA **6** Center for Computational Astrophysics, Flatiron Institute, Simons Foundation, 162 Fifth Avenue, New York, NY 10010, USA **7** Space Telescope Science Institute, 3700 San Martin Drive, Baltimore, MD 21218, USA ¶ Corresponding author

## Summary

We find ourselves on the brink of an exciting era in observational astrophysics, driven by groundbreaking facilities like JWST, Euclid, Rubin, Roman, SKA, or ELT. Simultaneously, computational astrophysics has shown significant strides, yielding highly realistic galaxy formation simulations, thanks to both hardware and software enhancements. Bridging the gap between simulations and observations has become paramount for meaningful comparisons.

We introduce py-ananke, a Python pipeline designed to generate synthetic resolved stellar surveys from cosmological simulations, adaptable to various instruments. Building upon its predecessor, ananke by Sanderson et al. (2020), which produced Gaia DR2 mock star surveys, the py-ananke package offers a user-friendly "plug & play" experience. The pipeline employs cutting-edge phase-space density estimation and initial mass function sampling to convert particle data into synthetic stars, while interpolating pre-computed stellar isochrone tracks for photometry. Additionally, it includes modules for estimating interstellar reddening, dust-induced extinctions, and for quantifying errors through dedicated modeling approaches. py-ananke promises to serve as a vital bridge between computational astrophysics and observational astronomy, facilitating preparations and making scientific predictions for the next generation of telescopes.

## Statement of need

The upcoming decade holds promise for groundbreaking discoveries, thanks to a multitude of recent and forthcoming observational facilities. The James Webb Space Telescope (Gardner et al., 2006), for instance, with its exceptional specifications, has already delved into early universe galaxies with unprecedented detail, revealing their rich diversity (Adams et al., 2023; Casey et al., 2023; Eisenstein et al., 2023; Ferreira et al., 2022; Finkelstein et al., 2023; Harikane et al., 2023). The recently launched Euclid Telescope (Laureijs et al., 2011) promises to shed light on the universe's accelerating expansion by surveying an immense number of galaxies (Euclid Collaboration et al., 2022). The Vera Rubin Observatory (Ivezić et al., 2019), with first light expected soon, will precisely map the Milky Way (MW) up to the virial radius and nearby galaxies, providing exceptional stellar astrometry data. Furthermore, the Nancy Grace Roman Space Telescope (Akeson et al., 2019), set to launch in the next couple of years, will offer a wide field of view for deep-sky near-infrared exploration, facilitating the study of resolved stellar populations in nearby galaxies (Dey et al., 2023; Han et al., 2023) and our own (Sanderson et al., 2024). However, these observatories will generate an unprecedented amount of raw data, necessitating community preparedness.

In parallel, a number of projects have emerged over the last decade in computational astrophysics, continuously surpassing hardware and software limits to simulate galaxy formation in a cosmological context realistically (see Crain & van de Voort, 2023 for a recent review). These simulations serve as invaluable test beds in anticipation of the next-generation telescope era, but also for our own models. However, translating these simulations into mock observables is challenging due to the representation of stellar populations as star particles, with each particle representing a total stellar mass between $10^3$ and $10^8$ times the mass of the Sun. To compare simulations with real data, one must break down these particles into individual stars consistently. Since the simulation resolution is not "one star particle per star" in the vast majority of these simulations, producing mock observables necessarily requires a series of assumptions that can have different effects on the final prediction.

This challenge was addressed by Sanderson et al. (2020) when producing a mock Gaia DR2 catalog from Milky-Way-mass simulated galaxies in the latte suite of FIRE simulations (Wetzel et al., 2016) using the so-called ananke pipeline. They used phase-space density estimation and initial mass function sampling to transform particle data into individual synthetic stars, retaining parent particle age and metallicity. Photometry was determined by interpolating pre-computed stellar isochrone tracks from the Padova database (Marigo et al., 2017) based on star mass, age, and metallicity. Additional post-processing included estimating interstellar reddening, per-band dust extinctions using metal-enriched gas distribution, and error quantification based on a model described by functions calibrated to (Gaia Collaboration et al., 2018) characterizations.

The ananke pipeline by Sanderson et al. (2020), though powerful, lacks user-friendliness and flexibility. It is challenging to integrate into other pipelines and expand beyond the Gaia photometric system. The development of py-ananke aims to make this framework more accessible to a wider community. By providing a self-contained and easily installable Python package, it streamlines the ananke pipeline, automating tasks previously requiring manual intervention. py-ananke also expands ananke's photometric system support and employs a modular implementation for future enhancements, promising a smoother upgrade path for users.

## py-ananke's framework

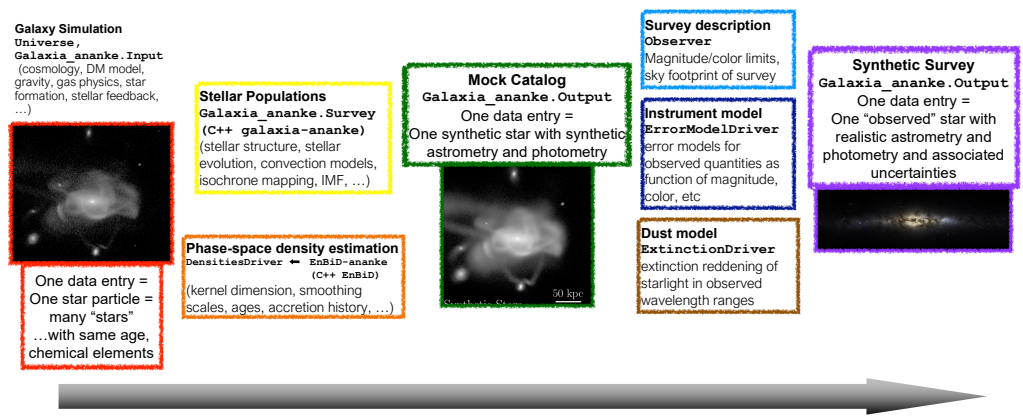

**Figure 1:** Schematic illustrating the inner framework of the `py-ananke` pipeline. The modules `py-EnBiD-ananke` and `py-Galaxia-ananke` are referred to by their import names `EnBiD_ananke` and `Galaxia_ananke`, with their respective `C++` backend softwares `EnBiD` and `galaxia-ananke`. The pipeline framework is illustrated from input to final output from left to right, showcasing the different objects and their purposes.

The implementation of py-ananke is designed to streamline the ananke pipeline, and to prevent the need for the user to manually handle the interface between `Python` and the `C++` backend software. It notably introduces dedicated wrapper submodules (hosted in repositories that are separate from that of py-ananke, but linked as `git` submodules), namely `py-EnBiD-ananke` and `py-Galaxia-ananke`, specifically developed to handle the installation and utilization of these `C++` subroutines, namely `EnBiD` (Sharma & Steinmetz, 2006) and a modified version of `Galaxia` (Sharma et al., 2011) called `galaxia-ananke`. Figure 1 illustrates the inner framework process of the full pipeline, showcasing the various module and submodule classes and where they are used in an input to output fashion from left to right.

## Past and Ongoing Applications

Sanderson et al. (2020)'s data have been in public use for 5 years and delivered on the promise of this technique, leading to the discovery of a new stellar stream (Necib et al., 2020), the development and validation of new machine learning methods for inferring the origins of stars (Ostdiek et al., 2020), insights into the formation history of the MW (Nikakhtar et al., 2021), searches for dark matter subhalos (Bazarov et al., 2022), and inference of the MW's interstellar dust distribution (Miller et al., 2022).

Likewise, a number of studies made use of Sanderson et al. (2020)'s existing ananke pipeline, often through the extensive effort to adapt it to other photometric systems:

- Shipp et al. (2023) investigated the detectability of MW stellar streams in the Dark Energy Survey (Abbott et al., 2018, 2021; Flaugher et al., 2015), for which they produced mock star catalogs mimicking DECam photometry from disrupted star clusters identified around simulated MW-mass galaxies
- Nguyen et al. (2024) produced a synthetic survey mimicking the third data release of Gaia (Gaia Collaboration et al., 2021, 2023), similar to how Sanderson et al. (2020) produced a synthetic survey of the second data release of Gaia (Gaia Collaboration et al., 2018)

These studies required significant effort caused by the challenges of using ananke, which py-ananke is designed to alleviate. Current ongoing projects are already using the new py-ananke package, and are benefiting significantly from its ergonomics.

# Acknowledgements

ACRT and RES acknowledge support from the Research Corporation through the Scialog Fellows program on Time Domain Astronomy, from National Science Foundation (NSF) grant AST-2007232, from NASA grant 19-ATP19-0068, and from HST-AR-15809 from the Space Telescope Science Institute (STScI), operated by AURA, Inc., under NASA contract NAS5-26555.

Package development used resources provided by the Frontera computing project at the Texas Advanced Computing Center (TACC). Frontera is made possible by NSF award OAC-1818253. Simulations used as test data for the package, and which form part of the example suite, were run using Early Science Allocation 1923870 and analyzed using computing resources supported by the Scientific Computing Core at the Flatiron Institute. This work used additional computational resources from the University of Texas at Austin and TACC, the NASA Advanced Supercomputing (NAS) Division and the NASA Center for Climate Simulation (NCCS), and the Extreme Science and Engineering Discovery Environment (XSEDE), which was supported by NSF grant number OCI-1053575.

Package development and testing was performed in part at the Aspen Center for Physics, supported by NSF grant PHY-1607611, and at the Kavli Institute for Theoretical Physics workshop "Dynamical Models for Stars and Gas in Galaxies in the Gaia Era" and 2019 Santa Barbara Gaia Sprint, supported in part by the NSF under Grant No. NSF PHY-1748958 and by the Heising-Simons Foundation.

The authors are grateful to Anthony Brown and Jos de Bruijne for their cooperation in building the Gaia error models. We also acknowledge the input and encouragement of the participants of the Gaia Sprints (2017–2019), Gaia Challenge series (2012-2019) and "anankethon" workshops. We also thank Josh Borrow and the other developers of `swiftsimio` (Borrow & Borrisov, 2020) for their help in preparing our documentation inspired by the latter.